\documentclass[aps,prb,superscriptaddress,amsmath,amssymb,showpacs]{revtex4}
\usepackage{amsmath,amssymb,color}
\usepackage{graphicx}
\usepackage{epstopdf}

\numberwithin{equation} {section}
\begin{document}
\title{\bf Deuteron Momentum Distribution in  KD$_2$PO$_4$} 
\author{G. Reiter}
\affiliation{Physics Department, University of Houston, 4800 Calhoun Road,
Houston, Texas 77204, USA}
\author{A. Shukla}
\affiliation{IMPMC, Universit«es Paris 6 et 7, CNRS, IPGP, 140 rue de Lourmel, 75015 Paris, France}
\author{P.M. Platzman}
\affiliation{Lucent Technology, 600 Mountain Ave, Murray Hill, New Jersey, 07974,USA}
\author{J. Mayers}
\affiliation{ISIS Facility, Rutherford Appleton Laboratory, Chilton, Didcot, Oxfordshire,
OX11 0QX, UK}

\date{\today}

\begin{abstract}
The momentum distribution in KD$_2$PO$_4$(DKDP) has been  measured using neutron Compton scattering above and below the weakly  first order paraelectric-ferroelectric phase transition(T=229K). There is very litte difference between the two distributions, and no sign of the coherence over two locations for the proton observed in the paraelectric phase, as in   KH$_2$PO$_4$(KDP). We conclude that the tunnel splitting must be much less than 20mev. The width of the distribution indicates that the effective potential for DKDP  is significantly softer than that for KDP. As electronic structure calculations indicate that the stiffness of the potential increases with the size of the coherent region locally undergoing soft mode fluctuations, we conclude  that there is a mass dependent quantum coherence length in both systems. 
\end{abstract}
\maketitle
The question of the nature of the paralectric-ferroelectric phase transition in KDP and related materials is still an open one, interpretations having historically oscillated back and forth from an order disorder transition\cite{slt} to one in which tunneling of the protons played a significant role\cite{blin}, and back again.\cite{koh} The interpretation that tunneling, or at least, the quantum nature of the proton,  played a role in the transition was reopened recently by the demonstration that one could see coherence over two sites in the momentum distribution of the protons\cite{rmp} in the paraelectric phase. Measurements of the momentum distribution for protons have only recently become available, and can be done for light ions, in particular deuterons, as well. It is therefore natural to ask whether one can observe coherence in the deuterated materials.   

 The momentum distribution of the protons or deuterons  is a direct reflection of the structure of their local environment, and in the most favorable circumstances, can be used to infer the Born-Oppenheimer potential(BO) for the motion of the proton.  It is known that the oxygen-oxygen separation in the deuterated material is significantly larger than  that in the protonated one, and that this has a significant effect on the BO potential that the ion sees.\cite{ubel, koh} The substitution of deuterons for protons is therefore not a simple isotopic change in a fixed potential, and it is therefore not possible to use the observed potential in KDP to predict what one should see in DKDP.  One expects from the increase in the O-O separation, that the deuteron BO potential will also be double well,  with a higher barrier and hence smaller tunnel splitting. The splitting in KDP was found to be 90 mev by inferring the potential from the shape of the momentum distribution. 

 The momentum distribution is measured using neutron Compton scattering(NCS). This   is inelastic neutron scattering in the limit of large momentum transfer, $\vec{q}$(30-100 $\AA^{-1})$, also called Deep Inelastic Neutron Scattering. In this limit,  the neutrons scatter from the individual deuterons in the same manner that freely moving  particles scatter from each other. The fraction of neutrons scattered into a given angle with a given energy depends  only on the probability that the deuteron had a particular momentum  at the time it was struck by the neutron,  n($\vec{p}$). There is scattering of the neutrons off of the other ions as well, with the center of the peak due to an ion of mass M located at an energy of $\frac{\hbar^{2}q^2}{2M}$  Due to the much heavier mass of the other ions in DKDP, this scattering is easily separated from that of the deuterons.  What is measured is then the impulse approximation limit of the usual neutron scattering function
 
\begin{equation}
\label{ysc}
 S_{IA}({\bf q},\omega)=\frac{ M}{q}J(\hat{q}, y)=\frac{ M}{ q}  \int n(\bf{p})~\delta ( y-{\vec p}{\cdot \hat {q})\;d{\bf p}}
\end{equation}
where $y=\frac{M}{q}\left(  \omega - { \frac{\hbar q^2}{2M}} \right)$ and $\hat{q}$ is a unit vector in the direction of the momentum transfer. 

The momentum distribution, $n({\bf p})$ can be extracted from the measurements in a manner described in detail in earlier work\cite{rs, rmn}. The Compton profile, J($\hat{q}$, y), is represented as a series expansion in Hermite polynomials and spherical harmonics.  Small corrections due to deviations from the impulse approximation are added, the total  is convolved with the instrumental resolution function, and the coefficients in the series expansion determined by a least squares fitting procedure. 

The measurements were done on the eVS instrument at ISIS(subsequently redesigned as Vesuvio), which is a time of flight, inverted geometry spectrometer. The measurements are made by rotating the crystal about an axis perpendicular to the the plane containing the  beam and the detectors, so that the vector $\hat{q}$ spans a plane in the crystal. More detail can be had by increasing the number of planes measured. In this case, because of the size of the crystal and the time available, only one plane of data was taken. In order to interpret the data, it is then necessary to assume that the momentum distribution is cylindrically symmetric about the hydrogen bond axis. This is not a strong approximation, as the only deviations from cylindrical symmetry about the bond axis are due to the interactions with surrounding ions, not the two oxygens forming the bond. While it was possible to see such deviations in the KDP measurements, they were not large, and inclusion of these terms did not greatly  affect the observed distribution along the bond. 

$n(p)$ is then given by the expansion 

\begin{equation}
 \label{inv}
 n(\vec{p}) = \frac{e^{-p'^2}} {(2\pi)^{\frac{3}{2}}\sigma_z \sigma_x^2}\sum\limits_{n,l}n!(-1)^n
a_{n,l} p^lL_n^{l+\frac{1}{2}}(p'^2) Y_{l0}(\hat p')
 \end{equation}

where the $ L_n^{l+\frac{1}{2}}(p'^2)$ are associated Laguerre polynomials. $\vec{p'}$ is related to $\vec{p}$ by\cite{exp}
\begin{equation}
 (p'_x, p'_y , p'_z)=(\frac{p_x }{\sqrt2 \sigma_x}, \frac{p_y }{\sqrt2 \sigma_x},\frac{p_z }{\sqrt2 \sigma_z}). 
 \end{equation}
 The parameters $\sigma_x, \sigma_z$ are scale factors and determine by themselves the kinetic energy when the terms with n=1, l=0 and l=2, n=0 are omitted\cite{braz}. In units in which the energy is measured in mev and the momentum in $\AA^{-1}$, K.E.=1.045(2$\sigma_x^2+\sigma_z^2$).  The $a_n$ are arbitrary coefficients to be determined by the least square fitting process, along with the scale factors.  $a_{0,0}=1$. This expansion is complete, and can represent any cylindrically symmetric function. We find that the only term that is statistically significant  in these experiments is the n=2, l=0 term. The measured parameters are given in Table 1.

 \begin{figure}[h] 
   \centering
   \includegraphics[width=5in]{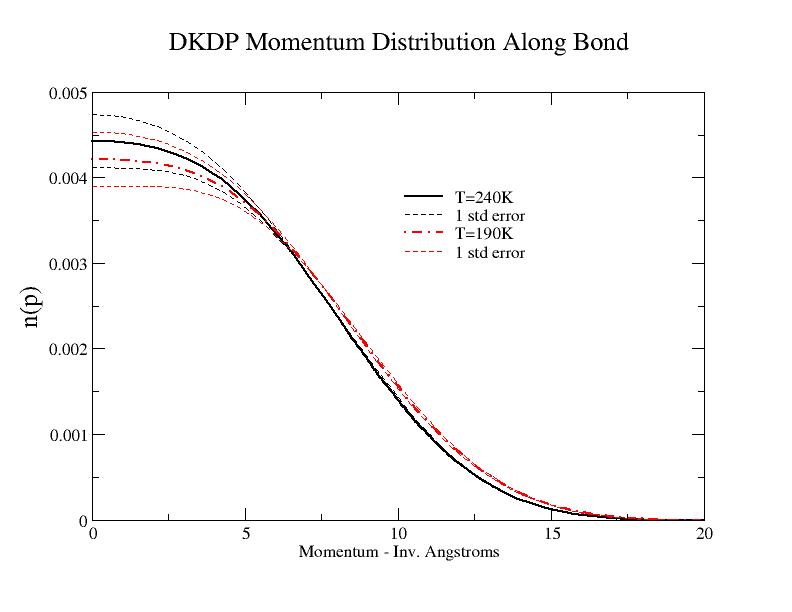} 
   \caption{The momentum distribution in DKDP along the bond, at temperatures above and below the ferroelectric-paraelectric phase transition. The results are nearly independent of temperature, and have been compared with a Gaussian fit corresponding to an harmonic well. }
   \label{fig1:data}
\end{figure}
We have included for later reference the parameters for KDP, which have not previously been published.  There are many more significant anharmonic parameters for KDP that have not been shown. We show in Figs. 1 and 2 the measured momentum distributions along the bond and transverse to the bond, repectively, for both temperatures, and compared with a simple Gaussian fit. 
 \begin{table}
\small
\caption{Parameters for Fit}
\centering
\begin{tabular}{||c|c|c|c||}
Temperature& $\sigma_x (\AA^{-1})$ &$\sigma_z(\AA^{-1})$& $a_{2,0}$ \\
\hline
190K&4.32&$5.58$&$-.246\pm .027$\\
\hline
 240K&4.33&$5.47$&$-.247\pm.037$\\
 \hline
 kdp-90K&4.14&5.69&.720$\pm .158$\\
 \hline
 kdp-130K&4.24&5.51&.189$\pm.048$\\
\end{tabular}
\end{table}

\begin{figure}[h] 
   \centering
   \includegraphics[width=5in]{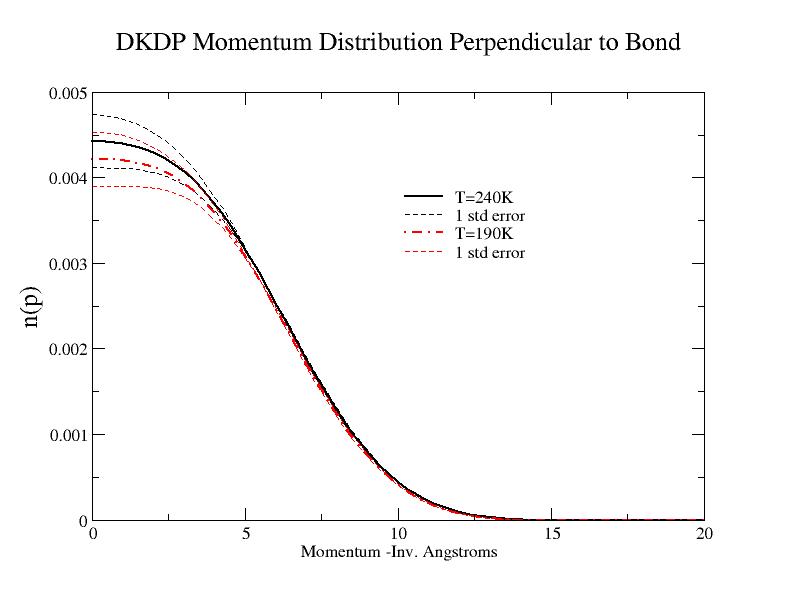} 
   \caption{The momentum distribution perpendicular to the bond for temperatures above and below the phase transition. Both perpendicular directions are assumed equivalent.  }
   \label{fig2:data}
\end{figure}
 \begin{figure}[h] 
   \centering
   \includegraphics[width=5in]{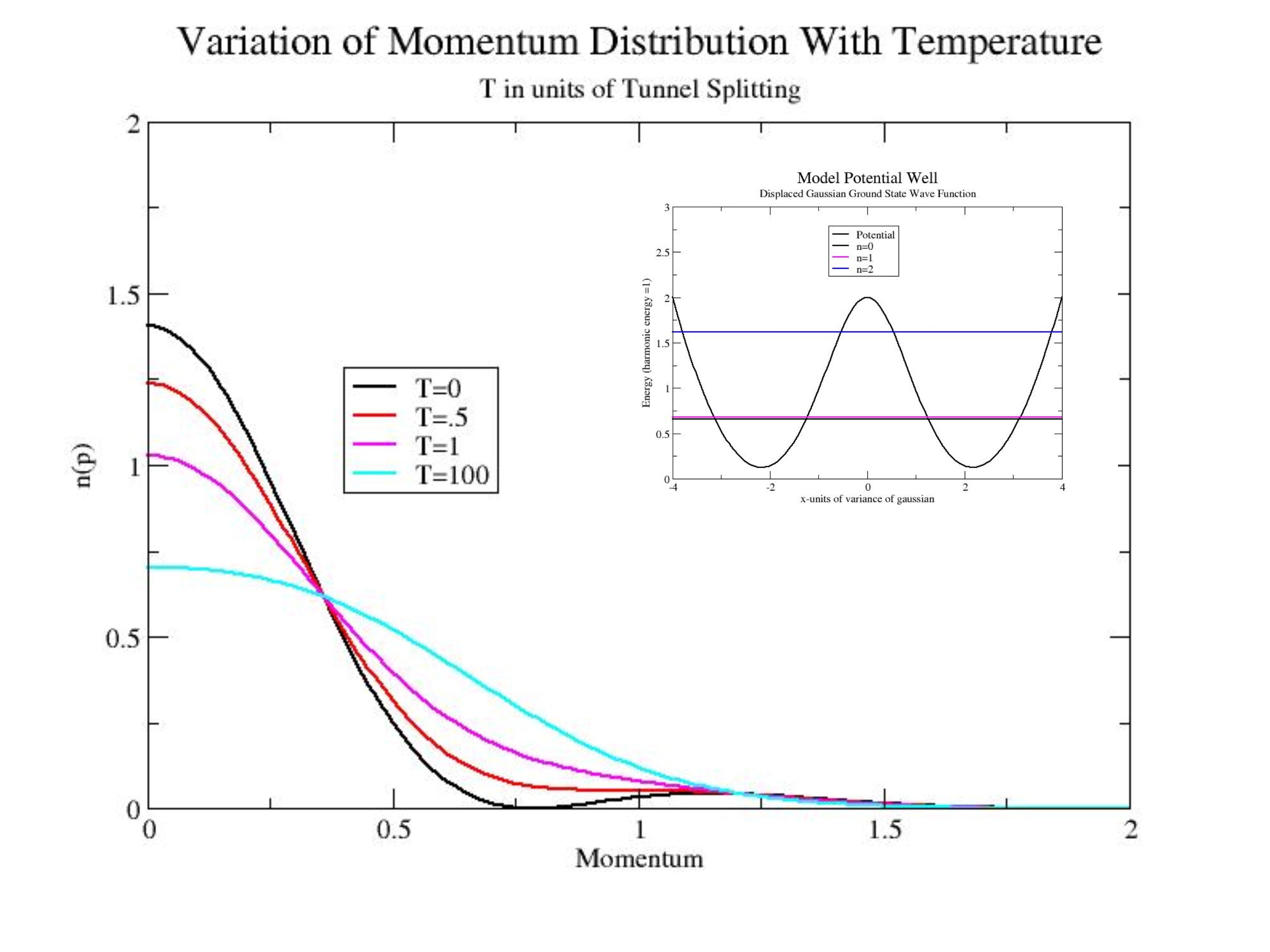} 
   \caption{The variation of the observed momentum distribution with the ratio of the temperature to the tunnel splitting(the energy difference between the ground state and first excited state)for a symmetric double well potential, shown in the insert.   }
   \label{fig3}
\end{figure}
 \begin{figure}[h] 
   \centering
   \includegraphics[width=5in]{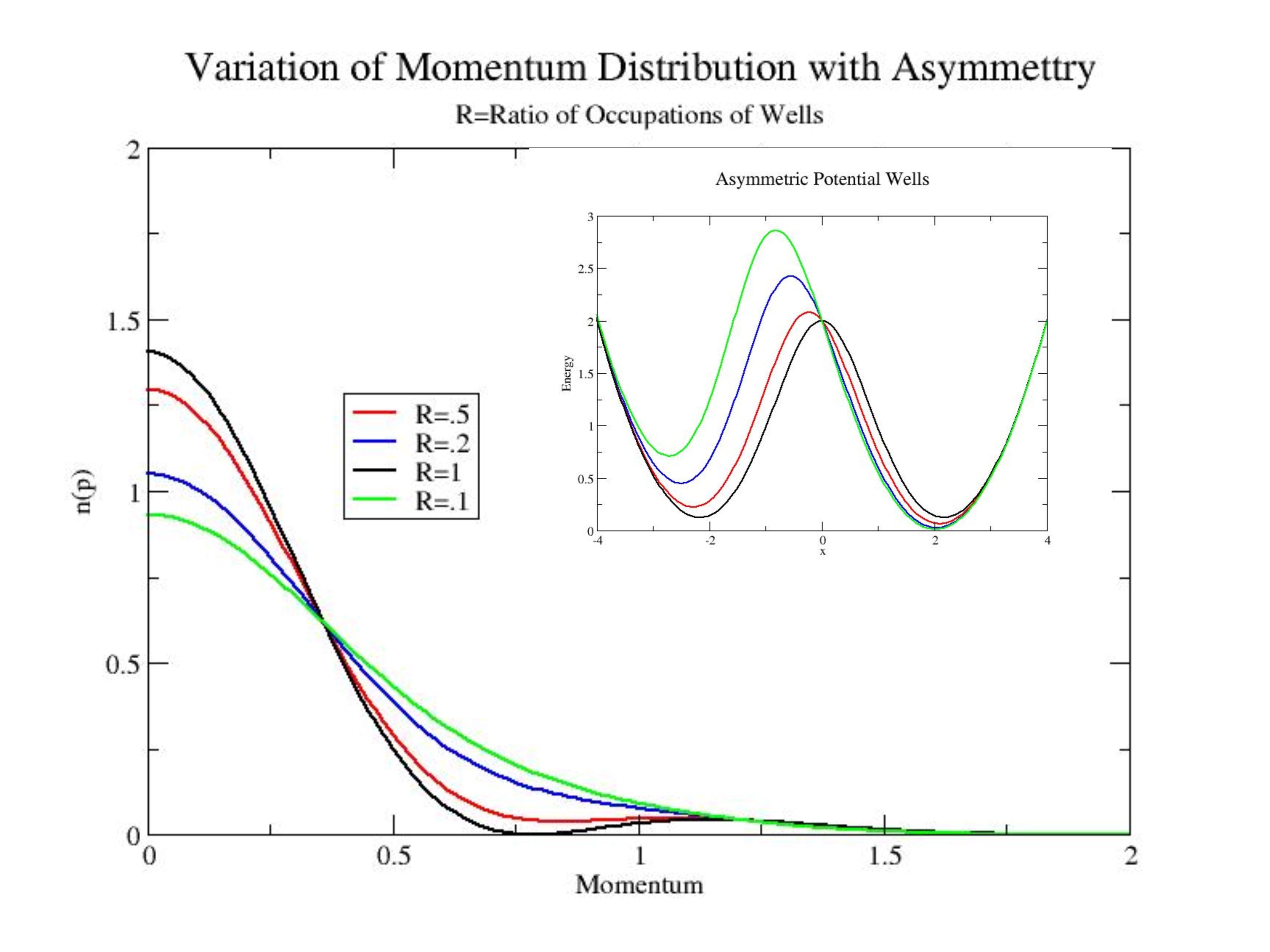} 
   \caption{The variation of the observed momentum distribution at T=0 for a well with an increasing asymmetry. R is the relative amplitude of the wave function in the shallow well compared to the deep one. }
   \label{fig4}
\end{figure}

It is immediately evident that there is very little change in the momentum distribution along the bond and less perpendicular to it, in going through the phase transition, and that the momentum distribution is nearly a Gaussian in both the bond and the perpendicular directions.  This was true for the transverse direction in KDP as well. Along the bond direction, however, the distribution  in KDP changed dramatically , being significantly narrower and showing  an oscillation characteristic of a double well potential in the paraelectric phase. \cite{rmp}  The lack of narrowing in DKDP along the bond can be interpreted in two ways. Either the deuteron is in  a symmetric double well, but the tunnel splitting is less than the temperature of the transition, or it is in an asymmetric double well above the transition, with equal populations in the two different asymmetric wells. We show  in Fig. 3 the variation of the momentum distribution for a simple model double well potential as the temperature varies. It is clear that  measurements must be in the limit of  kT$>>$$\Delta E$, where $\Delta E$ is the tunnel splitting, if no coherence is to be seen.

However, we see from Fig. 4 that an asymmetric double well behaves as a single well as far as the momentum distribution can tell when the ratio of occupations  in the ground state for the two sides of the well becomes$ <<$1. It is clear from the calculations of Koval et al\cite{koh}, and Zhang et al\cite{kio}, that below the transition, every well is asymmetric due to the coupling of the phosphorus motion to that of the protons.  We cannot decide from the qualitative nature of our data whether  this is true above Tc as well.  The small reduction in kinetic energy in the paraelectric phase seen in Fig. 1 could be due either to the remnants of coherence or the stiffening of the well as the system orders.   However, all other evidence points to the fact that the mechanism for the transition is the same in the deuterated and undeuterated systems.\cite{nelmes} When pressure is applied, so that the separation of the maxima in the wave function, $\delta$ in the two systems is the same, the Tc is the same. Since the evidence for coherence in KDP is clear\cite{rmp}, we conclude that the likelihood is that unobserved(because of the temperature) coherence in the deuteron motions is present in DKDP and the tunnel splitting is much less than 20mev, (in comparison to 90mev in KDP).

\newpage
The momentum distribution is very nearly a gaussian in both directions. This is true of KDP as well in the direction transverse to the bond, Assuming the potential in the transverse direction is harmonic out to distances that include the support of the wavefunction of the first excited state, we can calculate the excitation energy that would be observed in an infrared absorption or Raman measurement by fitting a Gaussian to the measured momentum distributions.  We obtain from the Gaussian fits  distribution frequencies of 700 cm$^-1$ for the transverse mode and 951 cm$^-1$ for the vibration along the bond. This compares well with values of the lowest two modes of the deuterated crystal obtained by Tominaga et al\cite{tom} from Raman scattering on mixed  KDP-DKDP crystal, of 721 cm$^{-1}$ and 973 cm$^{-1}$. However, they identify two modes of KDP at 1012 cm$^{-1}$ and 1312 cm$^{-1}$ with the equivalent  modes in KDP. While this is probably true for the lowest frequency  transverse mode, we think the identification of the higher frequency mode as an oscillation along the bond is erroneous, and its frequency scaling with the reduced mass fortuitous.    In the KDP case, we have two inequivalent directions perpendicular to the bond\cite{fel}, with excitation energies obtained as above from our own measurements of the transverse widths of 1015 cm$^-1$ and 1234 cm${^-1}$. This is to be compared with values of 1038 cm$^{-1}$ and 1293 cm$^{-1}$ obtained from infrared absorption measurements\cite{wei}.  While the frequency of the transverse modes in KDP and DKDP scales as m$^{\frac{1}{2}}$ (Tominaga et al use a reduced mass), as would be expected for a harmonic potential that was the same for both KDP and DKDP, the motion along the bond is far from harmonic in KDP, and the frequencies would not be expected to scale. We think the 1312 cm$^{-1}$ mode is actually the other transverse mode observed at 1293 cm$^{-1}$ , which is coincidentally in roughly the right place to be the analog in KDP of the mode at 973 cm$^{-1}$. There is further evidence in our data, the momentum distribution widths,  to indicate that the potentials along the bond in KDP and DKDP are quite different, in contadiction to the assumption of Tominaga et al. 

We emphasize, that the Figures 3 and 4 are based on a simple one particle in a potential well model of the system. Due to the interaction of the deuterons and the phosphorous, which the electronic structure calculations show to be essential for understanding the system in the paraelectric phase, the potential can only be thought of as an effective mean field potential, and the true nature of the transition may involve in an essential way the correlations in the motion of locally coupled ions. We think this is actually the case, because of the widths of the momentum distributions observed in DKDP. If the effective potential in DKDP were the same as that in KDP, the momentum widths $\sigma_x$ and $\sigma_z$  would be larger by a factor of 2$^{\frac 1 4}$ in DKDP. This change should be easily seen. This estimate is strictly true for a harmonic well, but the tendency to localize in space, and hence expand in momentum space as the mass increases is there for any well.  As table 1 shows,  $\sigma_z$ is actually smaller in DKDP than in KDP, and  $\sigma_x$, although larger, is not nearly as large as the simple scaling with mass would predict, indicating that there must be a compensating softening of the effective potential in DKDP. Just how to calculate this effective potential is an open question. Ab-initio calculations\cite{koh,kio} show clearly that the motion of the K and P  ions must be included with the motion of the protons to get realistic energy surfaces. Koval et al\cite{koh} study the effect of the simultaneous motion of clusters of sizes from 4 to 10 protons or deuterons along with the heavy particles, with the relative amplitude of the motion of the particles being given by that of the ferroelectric soft mode. They find  increasingly deep double well potentials for the collective coordinate, the soft mode amplitude, but in all cases the protons are in softer wells than the deuterons for the same size cluster.  We can reconcile their calculations with our measurements qualitatively if we assume that the size of the cluster that moves coherently above Tc,  presumably locally with amplitudes given by the soft mode, is smaller for the deuterated material than the protonated one. That is, that there is a mass dependent quantum coherence length in the problem that becomes smaller with increasing mass. As a consequence, there is a compensation of the greater depth of well for the deuterons in the  cluster of a given size  by a reduction in the size of the cluster for the deuterons relative to the protons. 

That the motion of all the surrounding protons or deuterons to a given set of phosphorous ions is strictly in phase with the soft mode amplitudes, as assumed in the theoretical calculations of Koval et al,  is probably not the case, but to the extent that coherent motion of the protons is playing a role, and their calculations indicate that it must to explain the depth of the effective potential we have measured in KDP, 
the decrease of the coherence length for coherent motion with mass is to be expected. The dependence of this length on temperature and pressure, and its role in the phase transition, remain to be explored.

\centerline{\bf Acknowledgements:}

This work  has been  supported by  DOE Grant DE-FG02-03ER46078.


\begin{thebibliography}{widest-label}
\bibitem{slt}J. C. Slater, J. Chem. Phys. {\bf 9}, 16 (1941)
\bibitem{blin}R. Blinc, J. Phys. Chem. Solids {\bf 13}, 204 (1960)
\bibitem{koh}S. Koval et al, Phys. Rev. B {\bf 71},184102 (2005) This paper contains a summary and references to the history of the problem. 
\bibitem{rmp}G. Reiter, J. Mayers  and P.  Platzman,  Phys. Rev. Letts. {\bf 89}, 135505  (2002)
\bibitem{ubel}J.M. Robertson and A .R. Ubbelohde, Proc. Royal Soc. of LondonA, {\bf  170}, 222 (1939)
\bibitem{rs}G. Reiter and R. Silver," Phys. Rev. Lett. 54, 1047 (1985).
\bibitem{rmn}G. Reiter, J. Mayers  and J. Noreland,   Phys. Rev.B {\bf 65}, 104305 (2002)
\bibitem{exp} This definition differs from that in Ref 4 in that a factor of $2^{2n+l}$ has been absorbed in the normalization of the Hermite polynomials in the expansion of the Compton profile.
\bibitem{braz}G. Reiter  et al. ,  Braz. J. Phys., {\bf 34}, 142 (2004) {see footnotes}
\bibitem{kio} Q. Zhang et al, Phys. Rev. B {\bf 65}, 024108 (2002)
\bibitem{nelmes}R. J.Nelmes, Ferroelectrics {\bf 124}, 355 (1991)
\bibitem{wei}E. Weiner, S. Levin and I. Pelah, J. Chem. Phys. {\bf 52}, 2881 (1970)
\bibitem{fel}J. Felcher and I. Pelah, J. Chem. Phys. {\bf 52}, 905, (1970)
\bibitem{tom} Y. Tominaga, Y. Kawahata and Y. Amo, Sol. St. Comm. {bf 125}, 419 (2003)
\end{thebibliography}
\end{document}